\documentclass{article}
\usepackage{ijcai24}

\usepackage{times}
\usepackage{soul}
\usepackage{url}
\usepackage[hidelinks]{hyperref}
\usepackage[utf8]{inputenc}
\usepackage[small]{caption}
\usepackage{graphicx}
\usepackage{amsmath}
\usepackage{amsthm}
\usepackage{mathbbol}
\usepackage{booktabs}
\usepackage{algorithm}
\usepackage{algorithmic}
\usepackage[switch]{lineno}
\usepackage{enumitem}

\usepackage{multirow}

\usepackage{tikz}
\usepackage[edges]{forest}
\usetikzlibrary{shapes, arrows.meta, positioning}
\usepackage{graphicx}
\usepackage{forest}
\usetikzlibrary{trees,positioning,shapes,shadows,arrows.meta}

\definecolor{myred}{RGB}{246, 204, 204}
\definecolor{myblue}{RGB}{164, 194, 244}
\definecolor{mygray}{RGB}{242, 242, 242}
\definecolor{myyellow}{RGB}{255, 221, 179}
\definecolor{myblueline}{RGB}{87, 127, 185}
\definecolor{bluelight1}{RGB}{185, 211, 237}
\definecolor{bluelight2}{RGB}{213, 222, 239}
\definecolor{mygreen}{RGB}{168, 209, 201}
\definecolor{greenlight}{RGB}{220, 235, 234}
\definecolor{hidden-draw}{RGB}{177, 177, 177}

\title{Data Augmentation for Sequential Recommendation: A Survey}

\author{
Yizhou Dang$^1$
\and
Enneng Yang$^1$\and
Yuting Liu$^1$ \and
Guibing Guo$^1$\thanks{Corresponding Authors.}\and
Linying Jiang$^1$\and
Jianzhe Zhao$^1$\and
Xingwei Wang$^{2\,*}$\\
\affiliations
$^1$Software College, Northeastern University, China\\
$^2$School of Computer Science and Engineering, Northeastern University, China\\
\emails
\{yizhoudang, ennengyang\}@stumail.neu.edu.cn,
\{guogb, jiangly, zhaojz\}@swc.neu.edu.cn,
wangxw@mail.neu.edu.cn
}

\begin{document}

\maketitle

\begin{abstract}
As an essential branch of recommender systems, sequential recommendation (SR) has received much attention due to its well-consistency with real-world situations. However, the widespread data sparsity issue limits the SR model’s performance. Therefore, researchers have proposed many data augmentation (DA) methods to mitigate this phenomenon and have achieved impressive progress. In this survey, we provide a comprehensive review of DA methods for SR. We start by introducing the research background and motivation. Then, we categorize existing methodologies regarding their augmentation principles, objects, and purposes. Next, we present a comparative discussion of their advantages and disadvantages, followed by the ‌exhibition and analysis of representative experimental results. Finally, we outline directions for future research and summarize this survey. We also maintain a repository with a paper list at \textcolor{blue}{ \url{https://github.com/KingGugu/DA-CL-4Rec}}.
\end{abstract}

\section{Introduction}
Sequential recommendation predicts future interactions by learning from users' historical sequence data. Over the last few years, many SR models based on different architectures \cite{tang2018personalized,kang2018self} have made significant achievements in modeling user behavior sequences. To take advantage of their complex architecture, an enormous amount of labeled training data is required. Unlike image and text data, which can be obtained through crowdsourced annotation or documentation, personalized recommendations rely on users' personalized behavior data \cite{yu2023self}. However, users tend to interact with only a few items on the platform. In addition, most of the historical data cannot be collected or used for model training due to cross-platform limitations and privacy protection  \cite{wu2021fedgnn}. Hence, the data sparsity problem significantly limits the recommendation performance of sequential models.

To tackle this issue, researchers realize that improvements in model structure may become challenging, so they turn their attention to DA. This method aims to mitigate data sparsity by broadening the diversity of original data or improving its quality without necessitating further data collection efforts \cite{ding2024data}. It has been widely used in computer vision (CV) \cite{yang2022image} and natural language processing (NLP) \cite{feng2021survey}. In SR, researchers have conducted many explorations and proposed many effective approaches. However, compared to a large amount of research at the model and algorithm level, the exploration of DA is still in the ascendant. With the rise of large language model (LLM) \cite{ding2024data} and the emphasis on data-centric AI \cite{lai2024survey}, large-quantity yet high-quality data is becoming vital for building a high-performance SR model.

\textbf{Contributions.} Given the growing interests and research, there is an urgent need for a timely survey to summarize the current achievements, discuss the strengths and limitations of existing research efforts, and promote future research. Therefore, this paper aims to (\romannumeral 1) present an up-to-date and comprehensive retrospective of DA methods for SR, (\romannumeral 2) provide comparative analysis and representative results between different augmentation approaches, and (\romannumeral 3) light on future directions to motivate and orient interest in this area. 

\textbf{Related Surveys.} Currently, several surveys focus on self-supervised learning covering some sequential DA methods \cite{ren2024comprehensive,jing2023contrastive,yu2023self}. However, the DA methods are only used as a means of constructing self-supervised signals in these surveys. They focus on how to perform self-supervised learning but neglect how to perform DA. Furthermore, many studies that directly use DA to improve model performance are not included. Our paper focuses on DA and provides a timely and comprehensive survey. Meanwhile, we also discuss a list of under-explored directions, which can shed great light on future research.

\textbf{Structure Overview.} The rest of this paper is organized as follows. In section \ref{Preliminaries}, we formulate the problem of SR, present what DA is, and how we use it in SR. After that, we introduce representative DA works for RS in section \ref{Heuristic-Based} and \ref{Model-Based}, in which we categorize existing methods based on their augmentation principles (heuristic-based and model-based) macroscopically. In section \ref{Discussion}, we further discuss the advantages and disadvantages of the different augmentation approaches and provide performance comparisons of representative methods. Finally, section \ref{Summary} lights on the future research directions and concludes this survey.

\begin{figure*}[!t]
    \centering
    \resizebox{\textwidth}{!}{
        \begin{forest}
            forked edges,
            for tree={
                grow=east,
                reversed=true,
                anchor=base west,
                parent anchor=east,
                child anchor=west,
                base=left,
                font=\small,
                rectangle,
                draw=hidden-draw,
                rounded corners,
                align=left,
                minimum width=4em,
                edge+={darkgray, line width=1pt},
                s sep=3pt,
                inner xsep=2pt,
                inner ysep=3pt,
                ver/.style={rotate=90, child anchor=north, parent anchor=south, anchor=center},
            },
            [
                {~Data Augmentation for Sequential Recommendation}
                , ver, 
                color=hidden-draw, fill=mygray!100, 
                text width=20em, 
                text=black
                [
                    Heuristic-based Augmentation (\ref{Heuristic-Based}), fill=myred!80, text width=13em, text=black
                    [
                    Data-level, fill=myred!80, text width=4em, text=black
                        [
                            Basic Data-level Operators (\ref{Basic}),  fill=myred!60,  text width=13.5em, text=black
                            [
                                \text{\cite{tang2018personalized,zhou2024contrastive}, \cite{xie2022contrastive}} \\ \text{\cite{sun2019bert4rec,tan2016improved}, \cite{liu2021contrastive}} \\
                                , color=hidden-draw, fill=myred!40,  text width=23em, text=black
                            ]
                        ]
                        [
                            Improved Data-level Operators (\ref{Improved}),  fill=myred!60, text width=13.5em, text=black
                            [
                                \text{Incorporating Side information: \cite{wang2022explanation},} \\
                                \text{\cite{tian2022temporal}, \cite{dang2023uniform,dang2023ticoserec},} \\
                                \text{\cite{tian2023periodicity}, \cite{tian2023periodicity}, \cite{xiao2024generic}} \\
                                \text{Scenario-specific Methods: \cite{oh2023muse}, \cite{li2023masked},} \\
                                \text{\cite{zhuang2024tau}, \cite{nian2024modeling}, \cite{dang2024repeated}} \\
                                , color=hidden-draw, fill=myred!40,  text width=23em, text=black
                            ]
                        ]
                    ]
                    [
                     Representation-level Operators (\ref{Representation}),  fill=myred!60, text width=14.5em, text=black
                        [
                            \text{\cite{qiu2022contrastive,du2023frequency}, \cite{ren2023contrastive},} \\
                            \text{\cite{xie2021adversarial}, \cite{chen2022intent,qin2024intent},} \\
                            \text{\cite{bian2022relevant}} \\
                            , color=hidden-draw, fill=myred!40,  text width=23.5em, text=black
                        ]
                    ]
                ]
                [
                   Model-based Augmentation (\ref{Model-Based}), fill=myblue!80, text width=13em, text=black
                       [
                             Sequence Extension and Refining (\ref{Prediction}), , fill=myblue!60, text width=14.5em, text=black
                            [
                            \text{Sequence Extension: \cite{liu2021augmenting}, \cite{jiang2021sequential},} \\
                            \text{\cite{wang2022learning}} \\
                            \text{Sequence Denoising and Refining: \cite{wang2021counterfactual}, } \\
                            \text{\cite{lin2023self}, \cite{zhang2024ssdrec}, \cite{yin2024dataset}} \\
                                , color=hidden-draw, fill=myblue!40, text width=23.5em, text=black
                           ] 
                       ]
                       [
                           Sequence Generation (\ref{Generation}), , fill=myblue!60, text width=14.5em, text=black
                           [
                            \text{Encoding-based Methods: \cite{wu2024personalized}, \cite{qin2023meta},} \\
                            \text{\cite{du2023contrastive}, \cite{hao2023learnable}, \cite{wang2022contrastvae}} \\
                            \text{Diffusion-based Methods: \cite{ma2024plug}, \cite{cui2024diffusion}} \\
                            \text{\cite{wu2023diff4rec}, \cite{liu2023diffusion}} \\
                                , color=hidden-draw, fill=myblue!40, text width=23.5em, text=black
                           ]
                       ]
                       [
                            LLM-Based Augmentation (\ref{LLM-Based}), , fill=myblue!60, text width=14.5em, text=black
                           [
                                \text{\cite{luo2024integrating}, \cite{wang2024large}, \cite{wang2024llm4dsr}}
                                , color=hidden-draw, fill=myblue!40, text width=23.5em, text=black
                           ] 
                       ]
                ]
            ]   
        \end{forest}
    }
\label{-35pt}
\vspace{-1em}
\caption{The taxonomy of data augmentation for sequential recommendation.}
\vspace{-1em}
\label{fig:taxonomy}
\end{figure*}
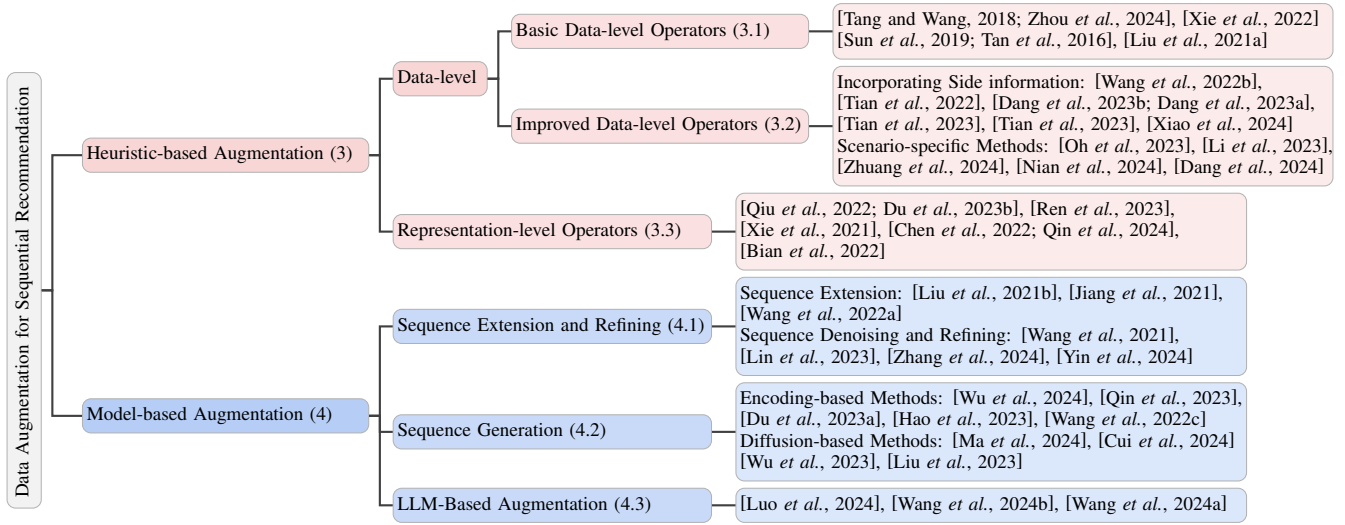

\begin{figure}[!t]
	\centering
	\includegraphics[scale=0.525]{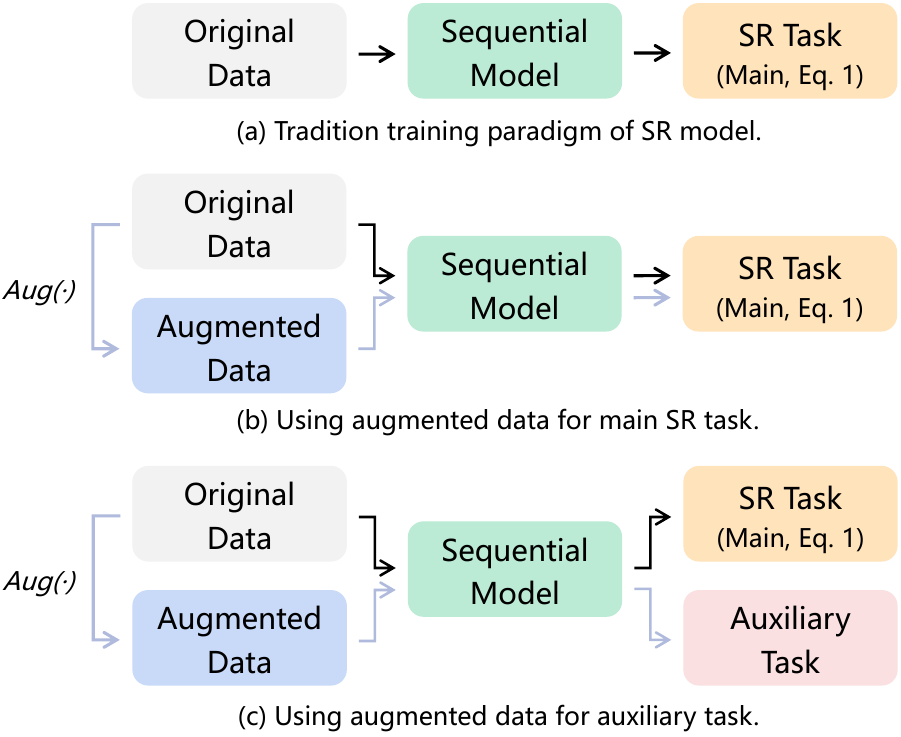}
        \vspace{-2mm}
	\caption{An illustration different training paradigms.}
	\label{fig:frame}
	\vspace{-2mm}
\end{figure}

\section{Preliminaries}\label{Preliminaries}
\subsection{Formulation of Sequential Recommendation}
Suppose we have user and item sets denoted by $\mathcal{U}$ and $\mathcal{V}$, respectively. Each user $u \in \mathcal{U}$ is associated with a sequence of interacted items in chronological order $s_u=[v_1, v_2, \ldots, v_{\left|s_u\right|}]$, where $v_j \in \mathcal{V}$ indicate the item that user $u$ has interacted with at time step $j$. The $\left|s_u\right|$ is the sequence length. Beyond the basic sequence data, we have auxiliary data denoted by $\mathcal{X}$. For example, multimodal data, attributes of items and users, social data, etc. Given the $s_u$ and $\mathcal{X}$, SR aims to accurately predict the possible item $v^{*}$ that user $u$ will interact with at time step $\left|s_u\right|+1$, formulated as follows:
\begin{equation}
    \underset{v^{*} \in \mathcal{V}}{\arg \max} \;\; P\left(v_{\left|s_u\right|+1}=v^{*} \mid s_u, \mathcal{X} \right).
\label{eq:SR}
\end{equation}
The SR model will calculate the probability of all candidate items and select the highest one for recommendation.

\subsection{What is Data Augmentation}
DA refers to increasing the size and diversity of the dataset without intentionally collecting or labeling more data. Existing methods obtain augmented data by performing transformations on existing data or directly synthesizing new data. Regarding Eq. \ref{eq:SR}, if algorithm and model improvements are focus on the process of calculating probability (i.e., $P(\cdot)$), then DA is focus on data $s_u$ and $\mathcal{X}$. We need to build an augmentation function $Aug(\cdot)$ to produce augmented data $s_u^\prime$ and $\mathcal{X}^\prime$. The primary purpose of DA is to improve the performance and generalization ability of machine learning (ML) models. It also can be considered a regularizer and mitigate overfitting problem when training ML models \cite{feng2021survey,hernandez2018data}.

\subsection{How Can We Use DA in SR}

As illustrated in Figure \ref{fig:frame}, in SR, the augmented data is usually used in two ways: for the main task of training the SR model (Figure \ref{fig:frame}(b)) or for auxiliary tasks in the training process (Figure \ref{fig:frame}(c)). The former directly improves the performance of the model. The latter, auxiliary tasks, are usually self-supervised learning or alignment, indirectly improving model representation learning or preference modeling capabilities. It is important to emphasize that in this survey, we focus on how to perform data augmentation, so how to build and perform auxiliary tasks is not part of our discussion.

\subsection{Exploration on Our Taxonomy}

We illustrate our taxonomy in Figure \ref{fig:taxonomy}. We categorize existing work into heuristic-based and model-based methods according to their principles macroscopically. \textbf{Heuristic-based} methods typically implement augmentation through randomized or heuristic data operators. They leverage randomness or observed patterns to perturb the original data to generate new data. Such methods usually do not require training, and the augmented data is known, given the original data and several hyperparameters. In this category, we further categorize the methods into data-level and representation-level based on their augmentation objects. \textbf{Model-based} methods typically train a data augmentation or data generation module, which usually contains learnable parameters. The augmented data is produced as a black-box process that follows a given goal or constraints. Since different approaches can be used to perform both data-level and model-level augmentation, we further categorize them according to augmentation purposes (e.g., sequence extension, refining, denoising, etc.).

\begin{figure}[!t]
	\centering
	\includegraphics[scale=0.4]{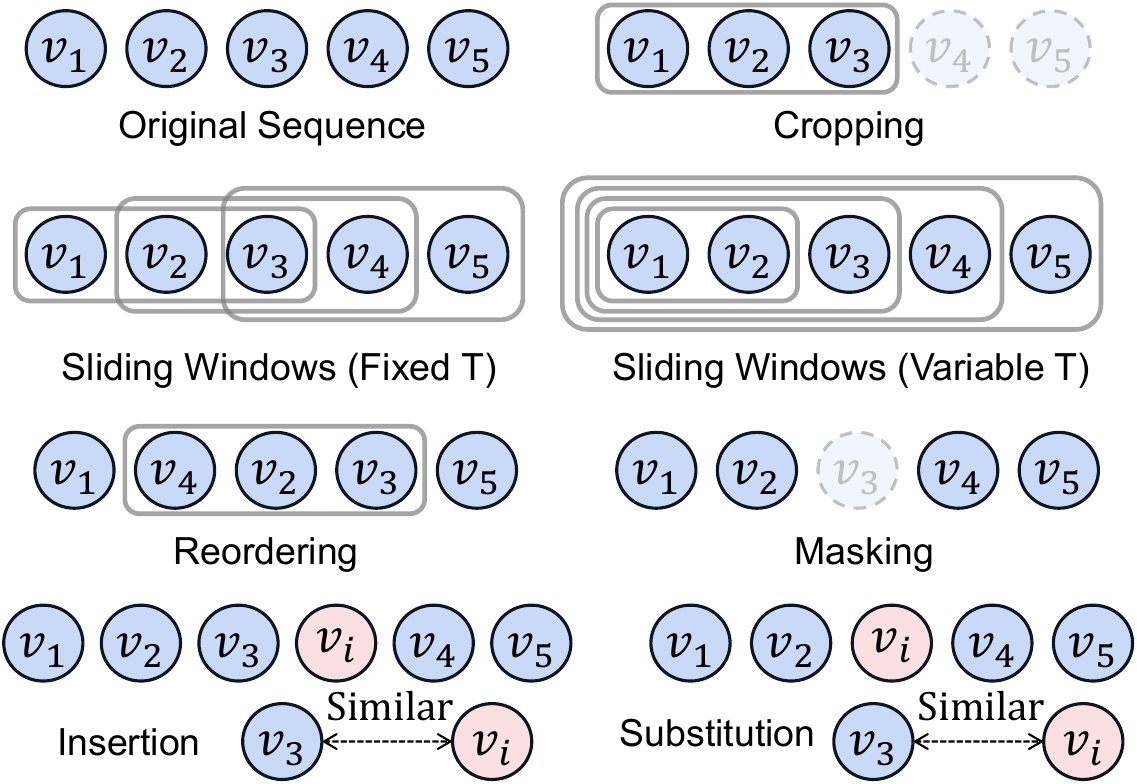}
        \vspace{-1mm}
	\caption{Basic data-level augmentation operations.}
	\label{fig:operator}
	\vspace{-2mm}
\end{figure}

\section{Heuristic-based Augmentation}\label{Heuristic-Based}

\subsection{Basic Data-level Operators}\label{Basic}

Data-level rule-based augmentation is the most widely used method in SR due to its simplicity. As illustrated in Figure \ref{fig:operator}, given an original sequence $s_u=[v_1, v_2, \ldots, v_{\left|s_u\right|}]$, the basic operators consist of seven classes as follow:

\textbf{Sliding Windows (SW):} Given a window length $T$ and $T < \left|s_u\right| $, this operation divides the original sequence into multiple sub-sequences by sliding a window from one end to the other \cite{tang2018personalized}:
\begin{equation}
\small
\{s_u^{a1},s_u^{a2},\cdot\cdot\cdot,s_u^{an}\} = Aug_{slide}(s_u).
\end{equation}
There can be several variants, including setting different or variable window lengths \cite{zhou2024contrastive}, step sizes, etc.

\textbf{Cropping:} Randomly select a continuous sub-sequence with length $L = \left|s_u\right| * \gamma$ from the original sequence. The $\gamma \in (0,1)$ is a hyper-parameter \cite{xie2022contrastive}.
\begin{equation}
\small
s_u^{a} = Aug_{crop}(s_u) = [v_i, v_{i+1}, \ldots, v_{i+L-1}].
\end{equation}
The Cropping can be seen as a simplified operation of Sliding Windows, as they all truncate the continuous sub-sequence from the original sequence.

\textbf{Reordering:} Randomly shuffle a continuous sub-sequence $\left[v_i, \cdots, v_{i+r-1}\right]$ of original sequence $s_u$ as $[v_i^{\prime}, \ldots, v_{i+r-1}^{\prime}]$. The calculation of sub-sequence length is the same as Cropping \cite{xie2022contrastive}:
\begin{equation}
\small
s_u^{a} = Aug_{reorder}(s_u) = [v_1, v_2, \cdots, v_i^{\prime}, \cdots, v_{i+r-1}^{\prime}, \cdots, v_{\left|s_u\right|}].
\end{equation}

\textbf{Masking:} Randomly mask a proportion $\eta$ ($0<\eta<1$) of items in the original sequence \cite{sun2019bert4rec}:
\begin{equation}
\small
s_u^{a} = Aug_{mask}(s_u) = [v_1^{\prime}, v_2^{\prime}, \ldots, v_{\left|s_u\right|}^{\prime}],
\end{equation}
where $v_i^{\prime}$ will be replaced with the `[mask]' token if $v_i$ is a selected item, otherwise $v_i^{\prime}=v_i$. The Masking can be regarded as the Dropout operation on sequence if we just delete the mask item \cite{tan2016improved}.

\textbf{Substitution:} Replace a proportion of items in the original sequence with correlated items \cite{liu2021contrastive}:
\begin{equation}
\small
s_u^{a} = Aug_{substitute}(s_u) = [v_1^{\prime}, v_2^{\prime}, \ldots, v_{\left|s_u\right|}^{\prime}],
\end{equation}
where $v_i^{\prime}$ will be replaced with the correlated item if $v_i$ is a selected item, otherwise $v_i^{\prime}=v_i$. The correlated item is obtained based on the correlation score or the similarity of item representation. 

\textbf{Insertion:} Insert a proportion of items into the original sequence \cite{liu2021contrastive}: 
\begin{equation}
\small
s_u^{a} = Aug_{insert}(s_u) = [v_1, \ldots, v_{id}, v_{id}^{\prime}, \ldots, v_{id}, v_{id}^{\prime}, \ldots, v_{\left|s_u\right|}],
\end{equation}
where $v_{id}^{\prime}$ is the item correlated to the one adjacent to the insertion position. Here, the way to obtain correlated items is similar to Substitution.

\subsection{Improved Data-level Operators}\label{Improved}
The augmentation process of basic operators involves excessive randomness, such as random selection of the operation position, sub-sequence length, and operation proportion. Excessive randomness may cause augmented data to lose key interactions, involve harmful noise, and have a semantic drift problem, making it difficult to improve performance or even impair performance. Therefore, many improved operators \cite{zhou2023equivariant,dang2023ticoserec} have been proposed.

\textbf{Incorporating Side information.} Side information (e.g., timestamp of interaction) acts as a navigation for the data augmentation operators, which indicates to the operator where and at what proportion to operate, thus improving the quality and diversity of the augmented data. A pioneer work, EC4SRec \cite{wang2022explanation} augmented sequences based on the importance of items determined by explanation methods, generating positive and negative views that better capture the underlying user intent and context. TCPSRec \cite{tian2022temporal} dividing user sequences into subsequences based on intervals to model the invariance and periodicity of user behaviors. Similarly, TiCoSeRec \cite{dang2023uniform,dang2023ticoserec} incorporated time information and proposed five interval-aware operators to transform non-uniform sequences into uniform ones, enhancing the model's ability to capture user preferences. Besides, EASE \cite{tian2023periodicity} generated periodicity-preserving sequences through item substitution and shuffling and creating periodicity-variant sequences by inserting virtual items or merging adjacent items to capture emanative periodicity better. A recent work, MBASR \cite{xiao2024generic}, considered multiple user behaviors (e.g., purchases and clicks) and proposed behavior-aware operators to generate diverse and informative training sequences. 

\textbf{Scenario-specific Methods.} In addition to introducing side information, some researchers propose targeted operators for sequences in specific recommendation scenarios. For example, MUSE \cite{oh2023muse} enriched shuffle play sessions in music RS by inserting frequently occurring transitions to reduce unique transitions, thereby enhancing the model's ability to capture and utilize sequential information. In addition to music recommendation, TAU \cite{zhuang2024tau} used uncertainty estimation to identify and complete the potentially missing check-ins in the user trajectory sequence. Besides, BTBR \cite{li2023masked} augmented basket sequence data by combining masking and swapping strategies, where item-level and basket-level masking are used to create diverse training samples, and an item-swapping strategy is applied to enrich interactions within baskets. FRec \cite{nian2024modeling} proposed a particular substitution operator that replaces specific items with the target item to generate explicit fatigue signals. These fatigue signals are further used to model user fatigue behavior. RepPad \cite{dang2024repeated} used the original sequence as the padding content instead of the special value `0' for widespread short sequences.

\subsection{Representation-level Operators}\label{Representation}
In addition to data augmentation operators at the data level, there are also some studies exploring data augmentation operators at the representation level. These methods operate within the feature or embedding space. For simplicity, we follow \cite{yu2023self} to denote the representation matrix with $\mathbf{R}$. Given $\mathbf{R}$, the augmentation operators include:

\textbf{Dropout:} Similar to deleting items in the sequence, this operator randomly drops a small portion of representations or embeddings \cite{qiu2022contrastive,du2023frequency}:
\begin{equation}
\small
\mathbf{R}_a=Aug_{dropout}(\mathbf{R})=\mathbf{R} \odot \mathbf{M},
\end{equation}
where $\mathbf{M}$ is the masking matrix that $\mathbf{M}_{i, j}=0$ if the $j$-th element of vector $i$ is masked/dropped, otherwise $\mathbf{M}_{i, j}=1$. The matrix $\mathbf{M}$ is generated by the Bernoulli distribution.

\textbf{Noise Injection:} In contrast to Dropout's discard, this operator injects uniform or Gaussian noise into the representation \cite{ren2023contrastive}. Here, we take Gaussian noise as an example:
\begin{equation}
\small
\mathbf{R}_a=Aug_{noise}(\mathbf{R})= \mathbf{R} + \epsilon \text{, where}\; \epsilon \in \mathcal{N}\left(0, \sigma^2 \mathrm{I}\right),
\end{equation}
where $\sigma$ is a hyperparameter to control the variance of Gaussian noise, the $\mathrm{I}$ denotes an identity matrix.

\textbf{Shuffling:} Randomly switches rows and columns in the representation matrix $\mathbf{R}$ to changing the contextual information, generating augmentations \cite{xie2021adversarial}.
\begin{equation}
\small
\mathbf{R}_a=Aug_{shuffle}(\mathbf{R})=\mathbf{P}_r \mathbf{R} \mathbf{P}_c,
\end{equation}
where $\mathbf{P}_r$ and $\mathbf{P}_c$ are permutation matrices that have exactly one entry of 1 in each row/column and 0 elsewhere.

\textbf{Clustering:} Clusters the representations by hypothesizing the presence of prototypes. The idea is that each user or item representation should closely align with these prototypes, which are identified through unsupervised learning methods such as the Expectation-Maximization (EM) algorithm \cite{chen2022intent,qin2024intent}:
\begin{equation}
\small
\tilde{\mathbf{C}}_a=Aug_{cluster}(\mathbf{R})=\operatorname{EM}(\mathbf{R}, \mathbf{C}),
\end{equation}
where $\mathbf{C}$ is the presupposed clusters (or prototypes) and $\tilde{\mathbf{C}}_a$ is the augmented prototype representations.

\textbf{Mixup:} Mixes the original user/item representations with representations from other users/items or previous versions to synthesize informative negative/positive examples \cite{bian2022relevant}. It usually interpolates two samples ($\mathbf{r}_i \in \mathbf{R}$, $\mathbf{r}_j \in \mathbf{R}$) in the following way:
\begin{equation}
\small
\tilde{\mathbf{r}}_i=Aug_{mixup}\left(\mathbf{r}_i\right)=\alpha \mathbf{r}_i+(1-\alpha) \mathbf{r}_j,
\end{equation}
where $\alpha \in[0,1]$ is the mixing coefficient that controls the proportion of information from $\mathbf{r}_i$.

\begin{figure}[!t]
	\centering
	\includegraphics[scale=0.365]{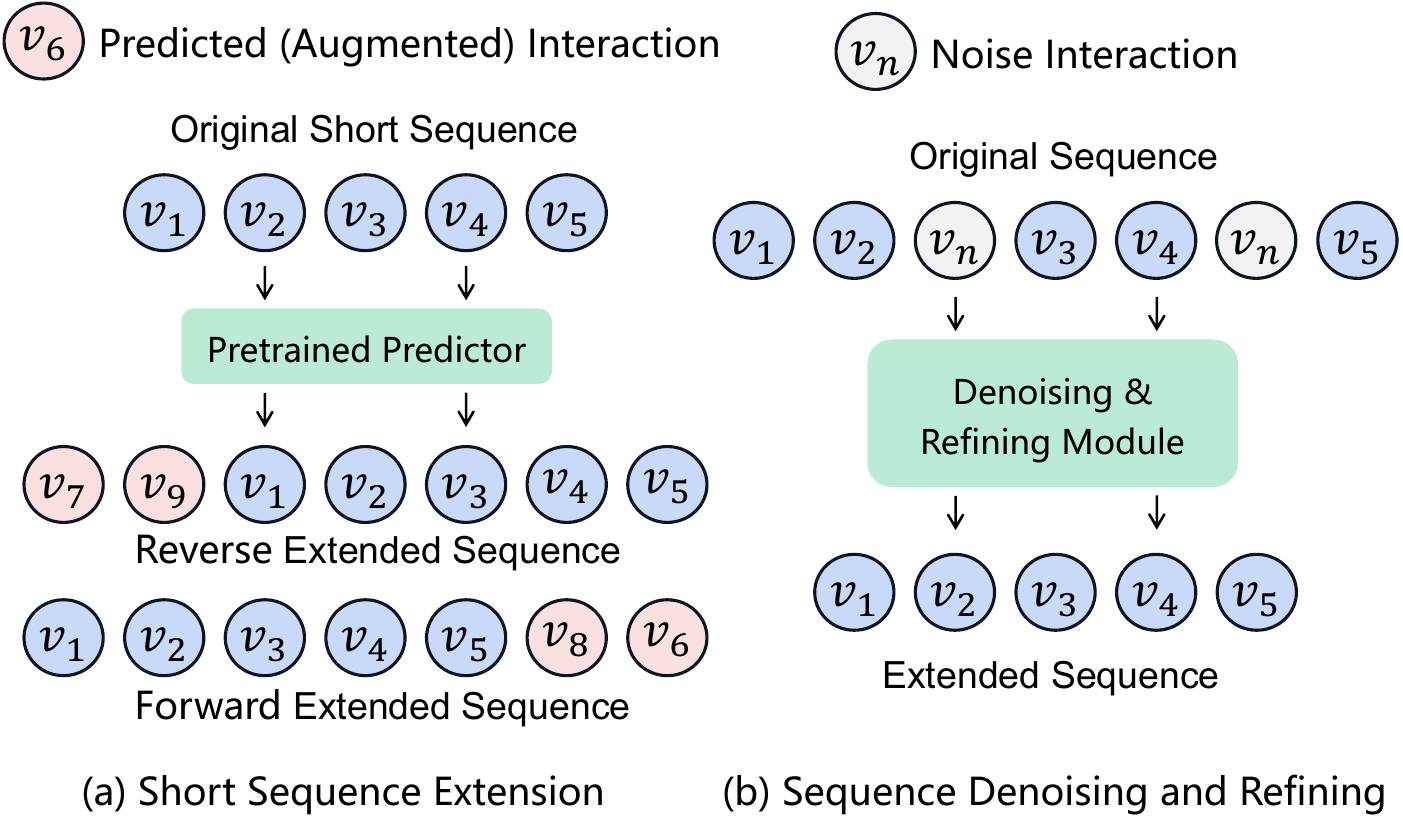}
        \vspace{-2mm}
	\caption{An example of sequence extension and refining methods.}
	\label{fig:prediction}
	\vspace{-2mm}
\end{figure}

\setlength{\tabcolsep}{1.25mm}{
\begin{table*}[!ht]
  \centering
  \scalebox{0.9}{
    \begin{tabular}{cc|cc}
    \toprule
    \multicolumn{2}{c|}{Heuristic-based Methods} & \multicolumn{2}{c}{Model-based Methods} \\
    \midrule
    Advantages  & Disadvantages  & Advantages  & Disadvantages \\
    \midrule
    Easy to deploy & Loss of key information & Multiple knowledge & Training and storage costs \\
    Training free & Conservative augmentation & Global perspective & Application restrictions \\
    Extreme sparse scenarios & Hyperparameter search & Multiple granularity & Extreme sparse scenarios \\
    Easy to customize & No external knowledge & Scalability \& Generalization & Process uncontrollability \\
    \bottomrule
    \end{tabular}}%
    \vspace{-2mm}
    \caption{Summary of the advantages and disadvantages of the two main types of data augmentation methods.}
    \vspace{-2mm}
  \label{tab:proscons}%
\end{table*}}%

\section{Model-based Augmentation}\label{Model-Based}

\subsection{Sequence Extension and Refining}\label{Prediction}
Methods in this branch typically train a prediction or decision module with additional learning objectives. These modules contain hyperparameters and learnable parameters. Given an original sequence, the prediction module can extend the sequence in the past or future direction. The decision module can select an appropriate augmentation operator or location for the original sequence, aiming to achieve denoising or adaptive augmentation of the original sequence. We illustrate two main types of methods in Figure \ref{fig:prediction}.

\textbf{Short Sequence Extension.} Due to the sparse user behavior, most users have short sequences, which directly limits the model's performance. To tackle this issue, ASReP \cite{liu2021augmenting} employed a reversely pre-trained transformer to generate pseudo-prior items for short sequences. Afterward, the augmented short sequence will be used to improve the performance of the model in predicting future interactions. Following this line, BARec \cite{jiang2021sequential} improved ASReP with a forward learning constraint to capture the contextual information when generating items reversely. Since the forward constraint is in line with the SR task, it can bridge the gap between reverse augmentation and forward recommendation. Besides, L2Aug \cite{wang2022learning} defines core and casual users, i.e., users with long and short sequences, respectively. It proposed to learn an augmentation policy that generates synthetic interaction sequences from core user data to mimic casual user behavior, thereby improving the recommendation system's performance for casual users without sacrificing the experience of core users.

\textbf{Sequence Refining and Denoising.} Since collected user sequences may contain noise or low-quality data, work in this branch focuses on how to utilize DA methods to denoise or refine data. A representative work, CASR \cite{wang2021counterfactual}, leveraged counterfactual thinking to create alternative user interaction histories based on the question, ``What would a user prefer if their past interactions were different?". Besides, SSDRec \cite{zhang2024ssdrec} introduces a three-stage process: first, it encodes multi-faceted inter-sequence relations to establish prior knowledge; second, it performs self-augmentation by inserting selected items into sequences to enhance them; and third, it applies hierarchical denoising to remove noise and refine the sequences. Similar to SSDRec, STEAM \cite{lin2023self} proposed a self-correcting mechanism that detects and corrects misclicked or missed items in user sequences, with a self-supervised task to enhance the quality of the training data. A recent work, DR4SR \cite{yin2024dataset}, introduced a data-level framework to refine the original sequence into a more informative and generalizable one. It introduced a pre-training task and a diversity-promoted regenerator to create diverse yet high-quality data.

\subsection{Sequence Generation}\label{Generation}
The previous subsection mainly focused on augmenting based on a single sequence. In contrast, generation methods focus on the distribution of the data as a whole. They generate new data by understanding and capturing the intrinsic distributional characteristics of the original data or by sampling in the learned latent space and generating new data points.

\textbf{Encoding-based Methods.} Methods in this branch typically train a specialized encoder that performs additional encoding or transformations on the original data to generate a new sequence representation. The structure of this encoder can be a simple MLP \cite{qin2023meta}, an AutoEncoder \cite{wang2022contrastvae}, or an encoder shared with the SR model \cite{du2023contrastive}. For example, MCLRec \cite{qin2023meta} integrated meta-optimized contrastive learning with both data-level operators and MLP-based model augmentation layer to generate more informative and diverse contrastive pairs. Similarly with this work, PPR \cite{wu2024personalized} used MLP to generate personalized prompts for user sequences to improve cold-start recommendations, and LMA4Rec \cite{hao2023learnable} introduced a learnable dropout layer to generate contrastive pairs. As for AutoEncoder, ContrastVAE \cite{wang2022contrastvae} leveraged it to produce variational augmentation to solve the inconsistency problem led by conventional data augmentation methods.

\textbf{Diffusion-based Methods.} Diffusion models generate new data samples by gradually adding noise to the data and then learning to reverse this diffusion process. DiffuASR \cite{liu2023diffusion} adopted the diffusion model to the item sequence generation. It proposed a sequential U-Net to capture the sequence information while predicting the added noise. Besides, two guide strategies are designed to control the DiffuASR to generate items that correspond more to the preference contained in the raw sequence. Following this work, DiffCLRec \cite{cui2024diffusion} and Diff4Rec \cite{wu2023diff4rec} use context-guided diffusion model and curriculum-scheduled diffusion model, respectively, to generate higher quality user interaction sequences. In addition to generating sequences, PDRec \cite{ma2024plug} proposed a plug-in diffusion model that generates user preferences for both observed and unobserved items. 

\subsection{LLM-based Augmentation}\label{LLM-Based}
The power capability and world knowledge of large language model (LLM) allow them to be used to alleviate the widespread problem of data sparsity \cite{ding2024data}. In the area of SR, some pioneering work has explored the use of LLM for data augmentation. These works mainly focus on prompt or instruction augmentation, which inputs original data into large models and specific instructions to generate augmented data. Since instructions are given by humans based on observations and understood by LLM, and the LLM performs the DA process, this type of approach is a combination of heuristic-based and model-based augmentation.

For example, Llama4Rec \cite{luo2024integrating} enriched the sequence of interacted items with additional items predicted by the LLMs. Specifically, it randomly samples a list of un-interacted items and adopts the prompt to ask the LLM to predict the item most likely to be preferred by the user. This predicted item is then randomly inserted into the user's sequence, resulting in an augmented sequence. This augmented data is then used to train a more powerful SR model. Similarly, Wang et al. \cite{wang2024large} employed LLMs to generate informative training signals for cold-start items in recommender systems. It proposed a pairwise comparison prompt to leverage LLMs to infer user preferences between pairs of cold-start items. Besides, LLM4DSR \cite{wang2024llm4dsr} used a self-supervised fine-tuning process to enable the LLMs to identify and replace noisy items, thereby enhancing the quality of training data.

\section{Discussion}\label{Discussion}

\subsection{Pros and Cons of Different Methods}\label{PandC}

In this section, we discuss the advantages and disadvantages from multiple perspectives of the two main types of methods, heuristic-based and model-based. We summarize them in Table \ref{tab:proscons}, corresponding to each point below.

\textbf{Heuristic-based Methods} use the randomness operation to impose changes on the original sequence or representation to obtain augmented data, yielding many advantages. Firstly, the simple logic inside the operators makes them very easy to implement and deploy. Secondly, these operators usually do not contain learnable parameters and do not increase the size of the model \cite{dang2024repeated}. The use of these operators does not depend on a specific model structure and does not have an impact on it. Thirdly, in some cases of cold starts and extreme scarcity of datasets, using these methods can quickly boost the total amount of data available for training. Lastly, these methods can be improved and adapted depending on the usage scenario and data. For example, the augment location is determined based on additional temporal information contained in the data \cite{dang2023uniform}, or the appropriate operator is selected for a particular scenario \cite{oh2023muse}.

However, there are also some drawbacks. Firstly, randomness may result in the loss of critical interactions. For example, the sub-sequence truncated by Cropping operators may not contain interactions that reflect the user's main preferences. Such sequences containing semantic drift problems may not improve or even impair the model's performance. Incorporating Side information to guide the augmentation process of the operator can alleviate the above problem \cite{dang2023ticoserec}. Secondly, these methods may also result in overly conservative data augmentation, where the augmented data is highly similar to the original data and cannot effectively improve model performance \cite{bian2022relevant}. Thirdly, these heuristic operators usually contain one to many hyperparameters. The impact of different hyperparameters on performance is sometimes noticeable for different datasets. When multiple operators are used simultaneously, the tuning and searching hyperparameters is time-consuming and consumes computational resources. Lastly, heuristics methods are challenging in producing fine-grained or adaptive augmentations \cite{qin2023meta}. Since the augmentation logic is pre-designed, the operators cannot adapt the augmentation process to the characteristics of the data or the model with learned knowledge.

\textbf{Model-based Methods} typically designs or trains a specialized learnable augmentation module that generates augmented data based on the characteristics and distribution of the original data. This line of work has many advantages. Firstly, since the augmentation module is usually jointly trained with the SR model or shares some components with the SR model, it can utilize the user preferences and knowledge already captured by the SR model when performing augmentation \cite{wang2022learning}. Secondly, in contrast to heuristic-based methods that typically target a single sequence for augmentation, model-based methods can learn to generate augmented data from the overall distribution of the original data, utilizing global information and preferences \cite{jiang2021sequential}. Some approaches can also utilize knowledge beyond the recommendation data. Thirdly, model-based augmentation can adaptively integrate data-level and representation-level augmentations to produce augmented data of varying granularity and personalization \cite{wang2022contrastvae}. Lastly, well-trained augmentation modules can be deployed without a time-consuming tuning process. It does not require iterative tuning of hyperparameters based on different datasets as heuristic-based methods do, saving time and computational resources.

However, there are also some drawbacks. Firstly, introducing these modules increases the model size and undoubtedly incurs additional training and storage costs. In some cases where the cost is limited or the size of the original model is small, this issue can be even more critical \cite{dang2024repeated}. Secondly, Some methods place restrictions on the structure of the original model. For example, ASReP \cite{liu2021augmenting} and BARec \cite{jiang2021sequential} can only be used on transformer-based SR models. Thirdly, in cases where the raw data is extremely sparse, these modules may become ineffective or unusable. This is because the raw data is insufficient to complete the training of the module. Lastly, the generation and augmentation process is low in controllability. In contrast to heuristic operators, only inputs and outputs are visible for model-based methods. Problem troubleshooting is complex if unanticipated augmented data is obtained.

\setlength{\tabcolsep}{1.25mm}{
\begin{table}[!t]
  \centering
    \scalebox{0.78}{
    \begin{tabular}{cc|cccccc}
    \toprule
    \multicolumn{2}{c|}{\multirow{2}[2]{*}{Method}} & \multicolumn{2}{c}{Beauty} & \multicolumn{2}{c}{Sports} & \multicolumn{2}{c}{Yelp} \\
    \multicolumn{2}{c|}{} & N@10 & H@10 & N@10 & H@10 & N@10 & H@10 \\
    \midrule
    \multicolumn{2}{c|}{Base Model (SASRec)} & 0.0338 & 0.0638 & 0.0177 & 0.0321 & 0.0135 & 0.0271 \\
    \midrule
    \multirow{6}[2]{*}{Sec 3.1} & SW & 0.0312 & 0.0599 & 0.0198 & 0.0366 & 0.0131 & 0.0275 \\
          & Cropping & 0.0270 & 0.0539 & 0.0164 & 0.0310 & 0.0145 & 0.0294 \\
          & Masking & 0.0302 & 0.0592 & 0.0171 & 0.0318 & 0.0140 & 0.0280 \\
          & Reordering & 0.0316 & 0.0609 & 0.0166 & 0.0305 & 0.0131 & 0.0274 \\
          & Substitution & 0.0319 & 0.0614 & 0.0169 & 0.0318 & 0.0141 & 0.0293 \\
          & Insertion & 0.0345 & 0.0640 & 0.0186 & 0.0360 & 0.0149 & 0.0309 \\
    \midrule
    \multirow{4}[2]{*}{Sec 3.1+} & CL4SRec  & 0.0304 & 0.0574 & 0.0192 & 0.0374 & 0.0187 & 0.0349 \\
          & CoSeRec & 0.0323 & 0.0622 & 0.0188 & 0.0348 & 0.0155 & 0.0320 \\
          & CL4SRec* & 0.0366 & 0.0686 & 0.0221 & 0.0412 & 0.0153 & 0.0325 \\
          & CoSeRec* & 0.0379 & 0.0701 & 0.0235 & 0.0428 & 0.0184 & 0.0362 \\
    \midrule
    \multirow{3}[2]{*}{Sec 3.2} & RepPad & 0.0390 & 0.0726 & 0.0212 & 0.0416 & 0.0177 & 0.0359 \\
          & TiCoSeRec* & 0.0401 & 0.0738 & 0.0269 & 0.0486 & 0.0203 & 0.0390 \\
          & ECL-SR* & 0.0434 & 0.0782 & 0.0284 & 0.0533 & 0.0190 & 0.0373 \\
    \midrule
    \multirow{5}[1]{*}{Sec 3.3} & Dropout & 0.0363 & 0.0675 & 0.0197 & 0.0352 & 0.0143 & 0.0293 \\
          & Noise & 0.0348 & 0.0651 & 0.0187 & 0.0341 & 0.0137 & 0.0286 \\
          & ICLRec* & 0.0407 & 0.0744 & 0.0238 & 0.0437 & 0.0205 & 0.0409 \\
          & DuoRec* & 0.0411 & 0.0756 & 0.0282 & 0.0498 & 0.0183 & 0.0376 \\
          & FEARec* & 0.0432 & 0.0803 & 0.0291 & 0.0510 & 0.0196 & 0.0395 \\
    \midrule
    \multirow{4}[1]{*}{Sec 4.1} & ASReP & 0.0351 & 0.0664 & 0.0189 & 0.0355 & 0.0146 & 0.0302 \\
          & BARec & 0.0403 & 0.0745 & 0.0202 & 0.0375 & 0.0161 & 0.0339 \\
          & STEAM & 0.0396 & 0.0737 & 0.0213 & 0.0384 & 0.0157 & 0.0334 \\
          & DR4SR & 0.0359 & 0.0668 & 0.0195 & 0.0362 & 0.0168 & 0.0351 \\
    \midrule
          & ContrastVAE & 0.0380 & 0.0699 & 0.0198 & 0.0367 & 0.0147 & 0.0298 \\
    \multirow{3}[1]{*}{Sec 4.2} & MCLRec & 0.0423 & 0.0774 & 0.0275 & 0.0493 & 0.0202 & 0.0406 \\
          & DiffuASR & 0.0372 & 0.0679 & 0.0183 & 0.0418 & 0.0169 & 0.0345 \\
          & PDRec & 0.0357 & 0.0670 & 0.0187 & 0.0336 & 0.0145 & 0.0286 \\
    \bottomrule
    \end{tabular}}%
    \vspace{-2mm}
    \caption{Performance comparison of different methods. The `3.1+' represents combinations of basic operators proposed in existing work. CL4SRec uses Cropping, Masking, and Reordering. CoSeRec uses all the operators in `3.1' except SW. For the Heuristic-based method (3.1, 3.1+, 3.2, 3.3), `*' represents data augmentation with auxiliary tasks, e.g., contrastive learning, and without `*' indicates that only data augmentation is used.}
    \vspace{-2mm}
  \label{tab:performance}%
\end{table}}%

\subsection{Experimental Comparison}

\textbf{Backbone and Dataset.} We adopt the representative SR model, SASRec \cite{kang2018self}, as the backbone model. It can be used as a backbone for almost any model-based methods. Following previous work \cite{liu2021contrastive,dang2024repeated}, We use three datasets: Beauty and Sports are obtained from Amazon \cite{Amazon} with user reviews of products. Yelp\footnote{\url{https://www.yelp.com/dataset}} is a business dataset, and we use the transaction records after January 1st, 2019. Users/items with fewer than five interactions are filtered out. We adopt the leave-one-out strategy to partition each user’s item sequence into training, validation, and test sets. 

\textbf{Implementation and Evaluation.} To ensure fair and reliable comparisons, we only compare the open source methods. We set the embedding size to 64 and the batch size to 256 for all methods. The maximum sequence length is set to 50. We carefully set and tune all other hyper-parameters of each method as reported in the papers. We conduct five runs and report the average results for all methods. The evaluation metrics include Hit Ratio@10 (H@10) and Normalized Discounted Cumulative Gain@10 (N@10). We rank the prediction over the whole item set rather than negative sampling, otherwise leading to biased discoveries \cite{zhou2024contrastive}. Generally, \emph{greater} values imply \emph{better} ranking accuracy.

\textbf{Experimental Results.} From the results in Table \ref{tab:performance}, we have the following observations: (1) Heuristic-based methods may compromise the integrity of the original data or introduce excessive noise, resulting in a decrease in performance compared to the original model, which is consistent with the disadvantage in the previous subsection. (2) For some heuristic-based methods, such as CoSeRec, CL4Rec, and DuoRec, heuristic augmentation and auxiliary tasks need to work together to improve model performance significantly. The auxiliary tasks can serve as constraints on the generated data or utilize self-supervised learning to discover preference knowledge \cite{dang2023ticoserec}. (3) In general, model-based approaches are superior to heuristic-based approaches. This is because the latter can generate fine-grained and adaptive augmentations from the overall distribution of the data. However, considering storage and computational costs, heuristic methods are also highly competitive. Simple heuristic-based methods can also achieve satisfactory performance in some cases, such as RepPad on Beauty and Sports datasets. (4) Model-based approaches do not always perform as well as they should, suggesting that the augmentation module may not be successful in capturing the overall distribution of the data or may generate inefficiently augmented data. In addition, no method achieved optimal performance in all datasets, suggesting that it is necessary to select appropriate enhancement methods for different datasets.

\section{Summary and Future Directions}\label{Summary}
In this paper, we provide a comprehensive survey about data augmentation methods for sequential recommendation. These methods are categorized based on their augmentation principles, objects, and purposes. After that, we provide a comparative discussion of their advantages and disadvantages. Moreover, we conduct extensive experiments on three publicly available datasets for a holistic evaluation among different methods. Despite the appreciable achievements, we note that several open problems remain to be solved, which are summarized as follows.

\begin{itemize}[leftmargin=*]

\item \textbf{Theoretical Foundations of Data Augmentation.} Currently, most of the DA methods in the SR are proposed based on experience (basic operators), observation (improved operators), experiments, or existing basic methods (sequence generation methods). These methods lack theoretical foundations. We can merely know experimentally that data augmentation methods improve (in some cases, impair) the recommendation performance of a model, but we cannot explain this result theoretically. If theoretical support can be provided, it would be beneficial for us to choose appropriate data augmentation methods or quantity of augmented data for different scenarios, models, and datasets \cite{yang2022image}. It will also contribute to proposing new methods and improving existing methods.

\item \textbf{The Evaluation of Augmented Data.} The quality of data has a significant effect on the performance and generalization ability of the model. Currently, the primary way to evaluate DA methods is to compare their ability to improve the performance of the original model. How to evaluate the quality of the augmented data is still an unexplored issue. In the field of CV and NLP, we can assess the quality of a synthetic image or a piece of augmented text data either artificially or using models. However, existing methods cannot directly assess the merits of augmented sequence data and representations in SR qualitatively or quantitatively. Evaluating the quality of the augmented data can help us better evaluate and improve DA methods.

\item \textbf{The Balance between Relevance and Diversity.} Relevance and diversity are two essential attributes of augmented data. Relevance means that the augmented data should have transition patterns similar to the original data to avoid semantics drift problems. Diversity means that the augmented data should contain sufficient variations to enable the model to explore more user preferences and improve its performance \cite{bian2022relevant}. However, existing methods often only accommodate one of two. Heuristic-based methods that utilize stochastic operations can produce more diverse data but are likely to lack relevance. Model-based approaches tend to produce conservatively augmented data that lack novelty. Thus, there exists much exploration space for balancing these two attributes.

\item \textbf{Automated and Generalizable Augmentation Methods.} From Table \ref{tab:performance}, we can observe that the optimal DA method varies under different datasets. In practice, researchers often need to manually select the appropriate augmentation methods based on a particular dataset or model, which is often time-consuming and laborious \cite{dang2023ticoserec}. In addition, trained model-based methods often need to be retrained when facing new datasets \cite{yin2024dataset}. Therefore, it is necessary to propose schemes that can adaptively select DA methods or to propose generalized DA methods that do not need to be trained from scratch.

\item \textbf{Data Augmentation with LLM.} LLM has demonstrated its powerful capabilities. LLM's world knowledge makes reference to data augmentation beyond just the original data. Meanwhile, it can simultaneously utilize the human experience in heuristic-based methods as well as the overall distribution and hidden representation of data in model-based methods. Existing methods tend to obtain augmented data by constructing simple instructions that do not utilize the full potential of LLM. Also, LLM can sometimes produce repetitive or generic outputs that may not add meaningful diversity to the dataset \cite{ding2024data}. Leveraging LLM for data augmentation is a promising direction.

\end{itemize}


\bibliographystyle{named}
\bibliography{ijcai24}

\end{document}